\documentclass[review,conference]{IEEEtran}

\usepackage{graphicx}

\usepackage{amsmath,amssymb,amsfonts}
\usepackage{algorithmic}

\usepackage{graphicx}
\usepackage{textcomp}
\usepackage{hyperref}
\usepackage{xcolor}
\usepackage{booktabs}
\usepackage{verbatim}
\usepackage{xspace}
\usepackage{subfig}
\usepackage{url}
\usepackage[square, numbers]{natbib}

\setcitestyle{author,open={[},close={]}} 

\usepackage{color}
\usepackage{soul}
\usepackage{tikz} 

\definecolor{lightgrey}{rgb}{0.50, 0.50, 0.50}
\sethlcolor{lightgrey}
\newcommand{\gray}[1]{{\color{gray}\textbf{#1}}\xspace}
\newcommand{\codebox}[1]{\texttt{\colorbox{lightgrey}{#1}}}

\definecolor{black}{rgb}{0.00, 0.00, 0.00}
\sethlcolor{black}
\newcommand{\codeboxb}[1]{\texttt{\colorbox{black}{#1}}}

\newcommand{\summarybox}[2]{
\begin{quote}
{\it #1:} #2
\end{quote}}

\newlength{\leftbarwidth}
\newlength{\leftbarsep}
\colorlet{leftbarcolor}{black!120}

\usepackage{framed}
\renewenvironment{leftbar}{%
    \MakeFramed {\advance \hsize -\width \FrameRestore }%
}{%
    \endMakeFramed
}

\renewenvironment{quote}%
{\begin{leftbar}\noindent\hspace{-\leftbarwidth}\xspace}%
{\end{leftbar}}%

\newcommand{\gh}{{GitHub}\xspace}

\newcommand{\eg}{e.g.\xspace}

\newcommand{\etal}{\emph{et al.}\xspace}

\def\BibTeX{{\rm B\kern-.05em{\sc i\kern-.025em b}\kern-.08em
    T\kern-.1667em\lower.7ex\hbox{E}\kern-.125emX}}

\begin{document}

\author{\IEEEauthorblockN{Kamel Alrashedy}
\IEEEauthorblockA{\textit{Georgia Institute of Technology} \\
\textit{School of Computer Science}\\
Atlanta, GA, USA \\
kalrashedy3@gatech.edu}
\and
\IEEEauthorblockN{Ahmed Binjahlan}
\IEEEauthorblockA{\textit{Georgia Institute of Technology} \\
\textit{School of Electrical and Computer Engineering}\\
Atlanta, GA, USA \\
iihmto@gatech.edu}
}

\title{How do Software Engineering Researchers Use GitHub? An Empirical Study of Artifacts \& Impact
}

\maketitle

\begin{abstract}
Millions of developers share their code on open-source platforms like \gh, which offer \emph{social coding} opportunities such as distributed collaboration and popularity-based ranking. Software engineering researchers have joined in as well, hosting their research artifacts (tools, replication package \& datasets) in repositories, an action often marked as part of the publication's contribution. Yet a decade after the first such paper-with-GitHub-link, little is known about the fate of such repositories in practice. Do research repositories ever gain the interest of the developer community, or other researchers? If so, how often and why (not)? Does effort invested on \gh pay off with research impact? In short: we ask whether and how authors engage in social coding related to their research. We conduct a broad empirical investigation of repositories from published work, starting with ten thousand papers in top SE research venues, hand-annotating their 3449 GitHub (and Zenodo) links, and studying 309 paper-related repositories in detail. We find a wide distribution in popularity and impact, some strongly correlated with publication venue. These were often heavily informed by the authors' investment in terms of timely responsiveness and upkeep, which was often remarkably subpar by \gh's standards, if not absent altogether. Yet we also offer hope: popular repositories often go hand-in-hand with well-cited papers and achieve broad impact. Our findings suggest the need to rethink the research incentives and reward structure around research products requiring such sustained contributions.

\end{abstract}

\section{Introduction}\label{intro}
In the span of a decade, \gh 
has significantly transformed the Open Source Software (OSS) ecosystem. It now hosts over $100$ million repositories and $31$ million users, mostly software developers. A key difference of GitHub compared to traditional OSS hosting platforms is its ambition to be a \emph{social coding} platform: it encourages collaboration and discussion among developers through channels including ``issues" and ``pull requests", and facilitates discovery of repositories through popularity metrics such as ``stars" and ``forks". Its massive popularity has drawn many kinds of contributors, including world leading tech companies (\eg Google's Tensorflow \citep{Abadi}, Apache's Hadoop \citep{Shvachko}), developers aiming to demonstrate their skills, instructors hosting course material, and Software Engineering (SE) researchers.

The latter are the subjects of this work: research publications in SE venues often study programmers in open-source ecosystem, and also publish their own repositories with code \& data. When they do, these repositories can be considered, and are often described as, a significant part of the work's contribution. Indeed, most conferences now encourage authors to submit research artifacts along with their papers, primarily to facilitate replicability. The motivations for, and community uses of, creating such artifacts (on any platform) are well-studied \citep{Timperley, Hermann, Shepperd}. But most of these artifacts are static, snapshots created at the time of publication.  Comparatively little is known about the use of OSS, which allows an artifact to continue to evolve after publication, often explicitly in response to social interactions. Hosting a repository on \gh comes with certain social expectations -- GitHub markets itself as a ``social coding" platform, after all, in terms of community interaction and involvement, and with it the potential of mass collaboration and impact. Meeting such expectations requires continued involvement from authors, possibly long after publication. Why do some authors choose to do so, and to what extent do they succeed?

This work conducts a detailed study of the ``promises and perils" of hosting research code on GitHub (GH). We study ca 3,500 GH links from 10K SE publications in recent years, of which 309 correspond to repositories explicitly tied to the paper's contributions 306 of which are still available). We find a diverse mixture of goals and success: about one third of repositories gain virtually no recognition, and two-thirds no interaction, while a few dozen others achieve widespread (including extra-academic) impact. Those that succeed often respond more pro-actively to community interaction than the rest, many of which fail to do so even when bugs are pointed out.
This variation sometimes tracks with publication venue: papers in journals such as TSE rarely involve repositories, and when they do, those are rarely popular. Most of these publications use GH as a platform to store replication scripts \& data. In other communities, such as MSR and ICSME, publications more often come with GH repositories that involve tools. Those fail as often as they succeed to reach a wider audience; those that do frequently accompany highly cited papers. Indeed, in recent years, the presence of a popular GH repository correlates increasingly strongly with citation count.

Our findings raise questions about the incentives for research authors to maintain their projects and sustain their contributions after publication. At least some level of community interest was present on nearly all repositories, especially right after publication, not uncommonly from other researchers who pointed out problems with the provided materials. Response to these interactions was often quite poor, with many authors simply never replying to even multiple issues, or only doing so weeks or months later. We suggest that our community's emphasis on research paper impact (i.e., citations) over community impact (especially measurable in, but not limited to \gh interactions) may be responsible, and make several recommendations to improve this situation. Yet  we also argue that this need not be a trade-off: successful repositories correlated strongly with both citations and broader adoption. Our findings thus call for a change in narrative surrounding the contribution value of (especially) tool-related SE publications.

\section{Methodology}\label{sec:methodology}

The goal of this work is to provide an initial empirical and unbiased exploration of this domain. To this end, the analysis in this paper is entirely data-driven: starting with the initial corpus, we answer a series of general research questions each informed by, and yielding subsequent, arrays of \emph{findings}, which we present throughout the paper.
The high-level statistics of each stage of our paper and link collection \& annotation are shown in Table \ref{table:venues-papers}.

\begin{table}[t]
  \begin{center}
    \caption{Summary statistics of papers, GitHub URLs and author-published repositories for our selected venues.}
    \label{table:venues-papers}
    \begin{tabular}{r|cccc}
      \toprule
      & \textbf{Indexed} & \textbf{PDFs online} & & \\
      \textbf{Venues} & \textbf{Papers} & \textbf{(\% of papers)} & \textbf{GitHub urls} & \textbf{GH repos} \\
      \midrule
      \multicolumn{5}{l}{\emph{(International) Conferences}} \\
      \midrule
    ICSE & 8,517 & 4,059~~(47.7\%) & 754 & 62 \\ 
    ASE & 2,578 & 1,560~~(60.5\%) & 674 & 56 \\ 
    ESEC/FSE & 2,058 & 1,366~~(66.4\%) & 515 & 47 \\ 
    MSR & 824 & 651~~(79.0\%) & 356 & 34 \\ 
    ICSME & 545 & 346~~(63.5\%) & 268 & 31 \\ 
    SANER & 475 & 251~~(52.8\%) & 336 & 24 \\ 
      \midrule
      \multicolumn{3}{l}{\emph{Journals}} \\
      \midrule
    TSE & 1,963 & 931~~(47.4\%) & 110 & 22 \\ 
    EMSE & 1,572 & 869~~(55.3\%) & 613 & 33 \\ 
    \bottomrule
    \end{tabular}
  \end{center}
\end{table}

\subsection{Paper Selection}
\label{sec:selection}
The main objective of this paper is to gain insight into source-code repositories published on GitHub by software engineering researchers, so all our papers were selected from publications in top Software Engineering conferences and journals. We selected eight highly rated venues, including the two top-rated SE conferences (ICSE, FSE) and journals (TSE, EMSE), as well as four additional (well-rated) SE conferences where GitHub repository use appeared to be particularly common (ASE, MSR, ICSME, and SANER).

\begin{figure}[t]
  \centering
 {\includegraphics[width=7cm]{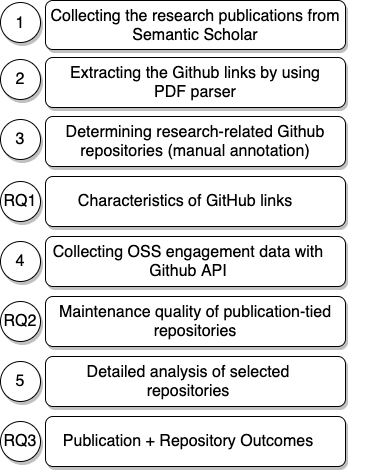}}
  \caption{Our data collection pipeline. We first collect ca. 20K papers from selected venues (1) and ca. 3.4K \gh links from these (2), which we manually annotate (3) to inform RQ1. We then gather the ``social" statistics of author-published repositories (4) for RQ2, and finally study some of these, especially more remarkable, repositories (and their papers) in-depth to gain richer insights for RQ3.}
  \label{fig:data_collection}
\end{figure}

\subsection{Data Collection}
We used the Semantic Scholar archive to find papers that were published at one of the selected top venues in software engineering (see Section \ref{sec:selection}). We collected the entire archive
and filtered it for papers with declared venues that match our selected options. The venue field is not standardized on Semantic Scholar, so we used heuristics to detect variations of these names (incl. abbreviations, partial abbreviations such as ``Intl. Conf.", etc.). This yielded 23,824 candidate papers. We next filtered out common confounding conference names (including many that resembled "International Conference on Software Engineering \underline{and} ...") and selected only those papers that had an attached URL to their PDF. Such PDFs are commonly hosted on ArXiv or on an author's personal webpage.

The frequency of such practice was rather different between venues, with some venues' PDFs online just about half the time,\footnote{This is a percentage of papers indexed by SemanticScholar, which likely under-counts the true number of papers.} while  publications at others, such as MSR, were released to the public over 80\% of the time -- a promising datapoint on the trend to open science. We note that the percentages for the older conferences are likely skewed by the greater proportion of their historical record preceding this practice; we found hundreds of papers from before the 2000s, but many thousands from the last decade (which is anyways more relevant to us, seeing as GitHub was introduced in 2008).
In all, this yielded 10,034 published papers with attached PDFs. We collected the following attributes for the resulting dataset, used in this study:
\emph{Publication date, Venue, Paper Link, Title, and Number of Citations}.

We next used \emph{PDFMiner}\footnote{https://pypi.org/project/pdfminer/} to extract all GitHub from these papers, the former (by simply matching any text containing \verb+GitHub.+). Such links were commonly found in footnotes or references.
In total, we found that $1,243$ publications with at least one GitHub link, with a total of $3,449$ such links between them. Circa half (613) of these papers had more than one link, and a small few (50) had over $10$ links. These links were manually annotated into three categories, as discussed in detail in Section \ref{sec:annotation}.
We additionally collected any links to Zenodo archives, which are sometimes set up to track the release of a GitHub repository. This creates a permanent archive of the code, which GitHub itself does not guarantee, and is therefore increasingly popular in the academic community. Since Zenodo does not expose an API to collect such data, we manually checked whether these linked archives (of which there were 107 across 77 papers, included in the above tallies) referenced a GitHub repository.

Finally, we used the GitHub GraphQL API to collect information about specifically those GitHub repositories that were created by the authors of the publications (as annotated in Section \ref{sec:annotation}). Most importantly, there are various ways to measure popularity of projects on GitHub. Primarily, users can ``star" projects that they are interested in. This is also a commonly used metric in scientific work to rank the popularity of repositories. It is also common to create a ``fork" of a project of interest, which allows making (potentially experimental) changes, and possibly submit these back to the original project in the form of ``Pull Requests", which thus reflect community interest in maintaining and improving a project. Finally, ``Issues" can be submitted to ask clarification, point out problems, and offer more general changes for discussion. These are often connected to a (later) Pull Request.
We thus collect the following properties through GitHub's API: \emph{Numbers of: stars, forks, commits, and counts \& properties of issues and pull requests (date open and closed, number of comments, author-involvement).} 

\subsection{Manual Annotation}
\label{sec:annotation}
We manually categorized all discovered links into three groups: 1. repositories created by the publication author(s) (of primary interest to this study), 2. repositories not owned by the paper's authors, 3. links that refers to non-repository pages on GitHub. The latter come in many variations, such as personal websites (on a `GitHub.io' address), organization or user profile pages, documentation or wiki pages, and links to specific issues, commits or pull-requests.
For the first category, we aim to identify GitHub repository that are proposed as part of the publication. These are often clearly identified, with links featured prominently in the abstract, introduction, or at the end of the paper, accompanied by phrases such as ``We make our tool/dataset publicly available at: ...". When in doubt, we opted to visit the GitHub profile associated with a repository (and/or its chief contributors) to identify whether the repository's owner is one of the publication's authors.

In total, category 1 contains slightly fewer papers than repository links; some papers contain multiple links, typically one for the paper's code and one for its dataset. In turn, but rarely so, links to the same repository may be featured in multiple papers, typically in the case of evolving research repositories associated with several publications. As may be expected, some papers referenced GH projects belonging to other publications, typically when using their tools. This was usually done through citations (\eg ``the tool from [x]"), rather than explicit links; only four (out of ca. 1,500) category 2 repository links explicitly referred to another paper's tool. Rather, these mostly referred to popular repositories on GH that were used for analysis or as anecdotal examples (further dissected in Section \ref{sec:broader}) -- for instance, Guava\footnote{\url{https://github.com/google/guava}} was featured in 10 papers.

\subsection{Objectives}
\label{sec:objectives}
We want to understand the nature of GitHub repositories released with published research. We do this in three steps, empirically studying a range of properties in each. First, we extract high-level characteristics of repository use in research by answering:\\
\textbf{RQ 1:} what characterizes GitHub links in published research? \emph{Sub-goals:}
\begin{itemize}
    \item \emph{1.1:} Identifying the purpose of GitHub links in publications.
    \item \emph{1.2:} Tracking changes in GitHub repository usage over time.
    \item \emph{1.3:} Breaking down repository inclusion by venue.
\end{itemize}

Next, we study the \emph{public interactions} and the associated maintenance burden of such repositories, focusing on whether and to what extent such social obligations are fulfilled after publication. \\
\textbf{RQ 2:} do repositories owners respond to public interactions? \emph{Sub-goals:}
\begin{itemize}
    \item \emph{2.1:} Characterizing the rate of common types of interaction.
    \item \emph{2.2:} Studying author responsiveness to public interactions.
    \item \emph{2.3:} Tracking trends in responsiveness over time.
\end{itemize}

Finally, we ask whether publishing code has a positive \emph{impact}, both as on the visibility \& usefulness of the research, and on the open-source domain. Here, we analyze correlations of various metrics, as well as examples in-depth. \\
\textbf{RQ 3:} Does Publishing Code add to Publication Impact? \emph{Sub-goals:}
\begin{itemize}
    \item \emph{3.1:} Analysis of whether releasing code, and its popularity, correlates with citation count.
    \item \emph{3.2:} Studying the extra-academic impact in OSS.
    \item \emph{3.3:} Finding references to GH-published tools in subsequent work.
\end{itemize}

\section{Results}
\label{sec:Results}
We now present the results to the research questions formulated in Section \ref{sec:objectives}. In each RQ-related sub-section, we present a series of empirical findings, which we reflect on in Section \ref{sec:discussion}.
We will release and archive our collected data upon paper acceptance.

\subsection{Characteristics of GitHub links}
We first explore the characteristics of the repositories connected to the 3,450 \gh links we annotated, studying their high-level purpose (in the referring paper), trends over time and publication venue, and various other statistics (such as programming language).

\subsubsection{The purpose of GitHub links in publications}
\label{sec:rq1_results}
\begin{figure}[t]
  \centering
 {\includegraphics[width=.8\linewidth,trim=0 0 0 10]{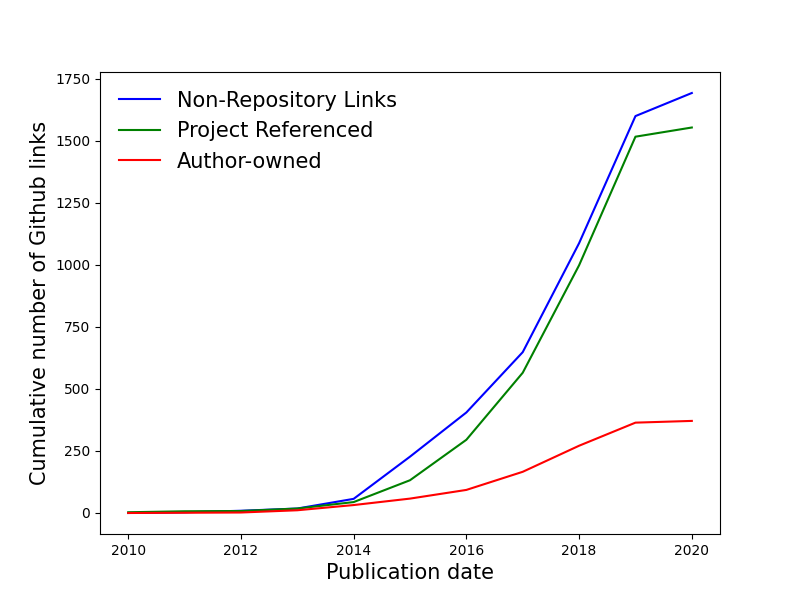}}
  \caption{Distribution and change of \gh links in selected publications over time.}
  \label{fig:link-usage}
\end{figure}

Our manual categorization of GitHub links in published research (see Section \ref{sec:annotation}) yields the high-level trends shown in Figure \ref{fig:link-usage}. In short, the vast majority of such references are not to repositories owned, or published, by the author, and references to \gh have increased steeply since 2016. The distribution of the types of reason for including such links has remained fairly stable over the years, with references to other projects and misc. \gh pages both ca. 5 times more common than author-owned repositories each. We discuss observed characteristics of the various categories here.

\textbf{Author-owned:} In total, we collected $309$ links to repositories owned and released by the paper's authors -- 11.2\% of the total number of links we annotated. Note that this necessarily excludes tools released in papers that were not accessible through our data collection process (e.g., whose PDF was not online, or indexed properly by SemanticScholar). It also does not include artifacts not hosted on \gh, because we specifically study social coding habits. It is thus significantly lower than the total number of artifacts released by our community, many of which were studied by \citep{Timperley}, who cite nearly 900 artifacts total across various platforms (but in turn did not study \gh repositories in as much depth as we do).

\summarybox{Finding \#1}{Only ca. one-third of authors choose to release artifacts specifically to \gh and reference it in their paper, suggesting a deliberate choice to cite it as a contribution.}

\begin{figure}[t]
  \centering
 {\includegraphics[width=.8\linewidth,trim=0 0 0 0]{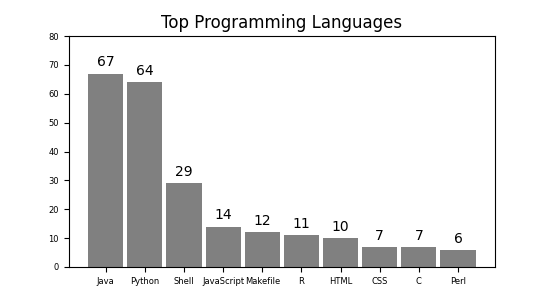}}
  \caption{The top $10$ programming languages used in author-owned repositories.}
  \label{fig:languages}
\end{figure}

The majority of these ``author-owned" repositories contained the implementations of tools that were proposed by the corresponding publication.
The other main use-case was to release replication data, including scripts and (small) datasets that played a significant role in producing the results in the paper.
Correspondingly the main programming language used in these repositories (shown in Figure \ref{fig:languages}) was often either Java, a popular choice for building tools, or Python, a common choice for one-off scripting, much like Shell (ranked third).\footnote{Python is also increasingly popular for building (often smaller) tools}.
In this paper, we are primarily interested in analyzing these repositories, which we do in more detail in Sections \ref{sec:rq2_results} and \ref{sec:rq3_results}

\summarybox{Finding \#2}{Authors use \gh repositories for a mix of purposes, including releasing general purpose tools, replication packages, and datasets -- the latter two of which generally do not require social coding features.}

\textbf{Projects Referenced:} recent publications mention a large number of GitHub repositories, regardless of ownership; we find $1,532$ of these. This is emblematic of recent research interest in studying data from OSS (both code \& communities): many of these references involved projects that authors used as data sources (\eg to train models) or examples of interactions in open-source communities. These links are spread irregularly over these papers: while $422$ publications refer to just one project, $111$ papers provide the remaining 1,098 links, and, among those, $14$ refer to $10$ or more. The latter frequently concerns comprehensive analyses and surveys of open-source projects.

\textbf{Misc. GH Links:} the remaining links point to other commonly used types of pages on the GitHub domain. This includes specific Pull-Requests ($136$), Issue pages ($183$) and Commits ($168$), as well as \verb+GitHub.io+ personal web-pages ($432$) and mentions of GitHub profile pages. The first three types of links are often used to demonstrate a motivating example, or proof of real-world impact, for a proposed tool. We note that this type of link was especially common in EMSE papers, which focus heavily on empirical work, while for instance being much less likely to appear in TSE papers.

\summarybox{Finding \#3}{\gh interest in publications extends well beyond releasing tools; in recent years the platform has become a go-to for many developer-centric insights,  especially in empirical studies.}

\subsubsection{Changes in GitHub repository use over time}
While \gh was launched in 2008, it naturally took several years before it was adopted by both developers and researchers.
Figure \ref{fig:link-usage} shows how this adoption of \gh in scientific work has changed since its launch in $2008$, focusing only on author-owned repositories now. It was rarely even mentioned in research publications until 2014, when a sudden uptake occurred, initially mainly in links to other repositories and pages, but soon also in repositories released by authors themselves. These curves follow, and slightly lag, GitHub's surge of popularity among developers in general, which commenced mid-2011 and broke 10M repositories by 2014.\footnote{\url{https://github.blog/2013-12-23-10-million-repositories/}}

\summarybox{Finding \#4}{\gh has become a common subject in SE research repositories starting in the mid 2010s, largely tracking its popularity among software developers.} 

On a positive note, virtually all of these repositories are still present: only $4$ out of the $309$ links we collected were unavailable. At the same time, GitHub's lack of archival status implies that many more of these artifacts are at the risk of disappearing. Usage of platforms such as Zenodo (which do archive permanently) to store GitHub repository releases was comparatively very thin in our dataset, and should perhaps be more actively encouraged.

\summarybox{Finding \#5}{Virtually all publication-tied repositories are still available at this time (2021), but the first handful of repository disappearances are a harbinger of the consequences of exclusively using a non-archival platform.}

\begin{figure}[t]
  \centering
 {\includegraphics[width=.8\linewidth,trim=0 0 0 10]{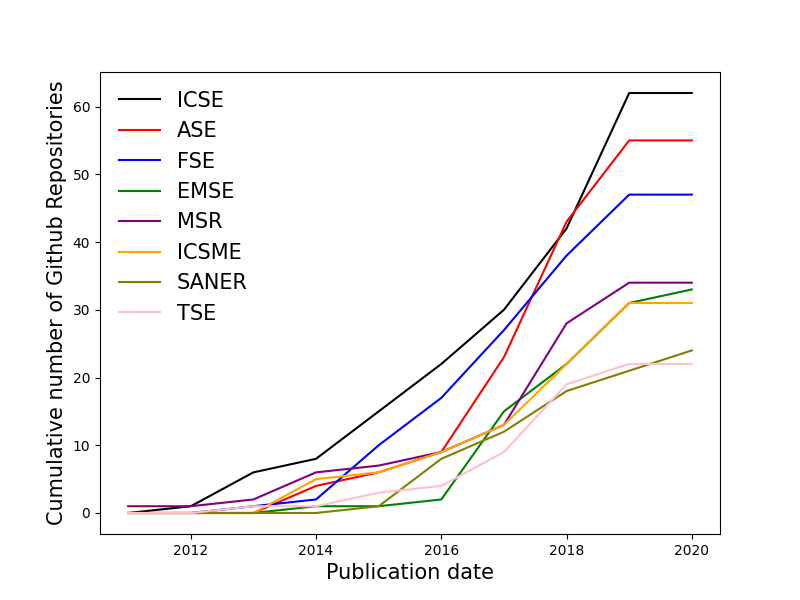}}
  \caption{Adoption of research-related GitHub repositories by venue over time. Note that recent (esp. 2020) tallies are partial, as papers and repositories may not yet be public.}
  \label{fig:use-change}
\end{figure}

\subsubsection{Repository inclusion by venue}

\begin{table*}[t]
  \centering
  \caption{Popularity statistics of author-owned GitHub repositories, by venue.}
  \label{tab:author-owned}
  \begin{tabular}{*{11}{c}}
 \hline
 & & \multicolumn{3}{c}{\textbf{Forks}} &\multicolumn{3}{c}{\textbf{Stars}} \\
 \cmidrule(lr){3-5} \cmidrule(lr){6-8} \cmidrule(lr){9-11}
  \textbf{Venue} & \textbf{Projects} & \textbf{Average} & \textbf{Max} & \textbf{Median} & \textbf{Average} & \textbf{Max} & \textbf{Median} \\ \hline
\textbf{ICSE}                 & 62                             & 18.0             & 328          & 3                      & 46.1             & 659          &         6.5       \\
\textbf{ASE}                  & 56                             & 23.8             & 254          & 3                         & 120.4            & 2105         & 8               \\
\textbf{FSE}                  & 47                             & 7.9              & 163          & 2                           & 22.5             & 218          & 6               \\
\textbf{MSR}                  & 34                             & 10.2              & 99           & 1.5                           & 38.5             & 483          & 4             \\
\textbf{ICSME}                & 31                             & 18.0              & 135          & 1.5                          & 28.8             & 512          & 5.5              \\
\textbf{SANER}                & 24                             & 9.5             & 68           & 1                          & 30.6             & 327          & 2               \\
\textbf{TSE}                  & 22                             & 1.9              & 11            & 1                            & 6.6              & 52           & 1               \\
\textbf{EMSE}                 & 33                             & 2.12              & 15            & 1                          & 6.0             & 54           &        2      \\

  \end{tabular}
\end{table*}

Finally, much in the way that some venues publish more empirical studies than others, we may expect differences in repository publication as well. Figure \ref{fig:use-change} breaks down this adoption of (specifically) author-owned GitHub repositories by publication venue in our field since 2014. Naturally, large conferences such as ICSE, FSE, and ASE have more publications that mention GH links, just by virtue of being larger. When accounting for the overall publication volume since 2014, nearly all venues published close to 10\% papers with associated GitHub repositories, with ICSE (5\%) and TSE (7\%) the lowest. In all cases, this trend has been upwards in recent years; for instance,  since 2018,14\% of all publications (and ca. 20\% of MSR, ASE, and ICSME papers) included a repository. Early number might suggest a slight slowdown since (from 14.7\% in 2018 to 13.7\% in 2019, not statistically significant).

\summarybox{Finding \#6}{\gh adoption has been fairly universal across publication venues and has largely increased as a share of papers in recent years.}

While adoption may be universal, popularity-related outcomes are far from egalitarian.
Table \ref{tab:author-owned} shows the distribution of various popularity related metrics on these repositories by publication venue. Overall, larger venues with many repositories (ICSE, ASE, and FSE, in terms of absolute counts) also tend to be fairly popular on average, as evidenced by e.g. a median of 6+ stars -- a humble, but non-negligible count, considering that half are (much) more popular.
The two journals (EMSE \& TSE), on the other hand, were remarkably unpopular, with just a few stars and forks on average, and few-to-none very popular outliers. We will analyze reasons for both these phenomena in greater detail in Section \ref{sec:maintenance}.
In all cases, the averages are much higher than the medians, signaling a strong skew in popularity. The most popular repository, `SecGen',\footnote{\url{https://github.com/cliffe/SecGen}, designed to introduce random insecurities into virtual machines and tied to a publication in ASE 2017 (among others).} had reached ca 2K stars and 300+ forks at the time of this writing. In most venues, highly popular outliers were about a hundred times more popular than the median, and vice versa, about a third of repositories had no forks and 1 or fewer stars.

\begin{table*}[t]
  \centering
  \caption{Popularity statistics of publication-associated GitHub repositories, by repository category.}
  \label{tab:author-owned_cat}
  \begin{tabular}{rccccccc}
 \hline
 & & \multicolumn{3}{c}{\textbf{Forks}} &\multicolumn{3}{c}{\textbf{Stars}} \\
 \cmidrule(lr){3-5} \cmidrule(lr){6-8} 
  \textbf{Category} & \textbf{Projects} & \textbf{Avg.} & \textbf{Max} & \textbf{Median} & \textbf{Avg.} & \textbf{Max} & \textbf{Median} \\ \hline
\textbf{Tool}                  & 236                             & 13.3              & 328          & 2                     & 53.5            & 2,105          & 6               \\
\textbf{Dataset}                 & 45                             & 15.7            & 261          & 1                  & 26.4             & 609          & 2               \\
\textbf{Rep. Package}                  & 18                             & 0.7             & 3          & 0                & 1.11            & 4         & 1               \\
\textbf{Tool \& Dataset}                  & 11                             & 2.5              & 17           & 0                        & 21.4             & 218          & 1             \\

  \end{tabular}
\end{table*}

We further explored the role of \emph{purpose}: some of these studied repositories evidently (and occasionally explicitly) serve as a data store, with minimal to no code included. Others are marketed as ``tools" or ``replication packages". We may reasonably expect a very different community response to these. We thus hand-annotated repositories based on their declared purpose (as per the README), identifying three main categories (tool, dataset, and replication package) and one additional group for the significant overlap between tool and dataset releases.\footnote{While the latter may be reminiscent of a replication package, it is notably different in that it does not included detailed steps to replicated the results of the paper, but rather ships a dataset along with the tool as a base requirement for the latter or an example to help users ``get started".}

Table \ref{tab:author-owned_cat} shows the frequency and popularity distribution of our repositories by category. Tools span the majority ($76.3\%$) of these, and are also the most popular category, with an average of $53.5$ stars (median: $6$+), evidencing that a large number of these tools are popular among GitHub users. After this, most remaining repositories belong to the dataset category or intersect between tools and datasets; the former rank especially high in terms of forks ($15.7$), perhaps because this is a convenient way to ``copy" the data. These also include a few outliers in terms of popularity that have attained broader appeal, such as Loghub.\footnote{\url{https://github.com/logpai/loghub}} The final category, containing replication packages, is comparatively utterly unpopular, with a median of 0 forks and 1 star. These are evidently not used by the broader open-source community, and would likely be better suited for an archival platform (see Section \ref{sec:discussion}).

\summarybox{Finding \#7}{Popularity varies tremendously among repositories, reminiscent of ``rich-get-richer" effects on other social platforms. This outcome is surprisingly often related with publication venue and repository purpose.}

\subsection{Repository Interactions and Maintenance}
\label{sec:rq2_results}
Publishing research artifacts on GitHub uniquely comes with the opportunity, and expectation, of communication with the public, and often the sustained upkeep that comes naturally with public response: GitHub is a social coding platform, on which unaffiliated developers are prone to reach out and discuss (and criticize) the submitted code. In this section, we study how well authors meet these expectations, and the ramifications for succeeding and failing to do so.

\begin{table}[t]
\centering
\caption{Mean public interactions with GitHub repositories, grouped by venue and overall, in terms of issues and pull requests created and closed. The corresponding resolution rate (``Rate") is an indicator of a repository owner's responsiveness to public interest.}
\label{tab:total_interactions}
\begin{tabular}{r|cccccc}
\textbf{Venue} & \multicolumn{3}{c|}{\textbf{Pull Requests}} & \multicolumn{3}{c}{\textbf{Issues}} \\ \hline
\textbf{} & \textbf{Created} & \textbf{Closed} & \textbf{Rate} & \textbf{Created} & \textbf{Closed} & \textbf{Rate} \\
\toprule
\textbf{ICSE} & 7.4 & 6.9 & 93.2\% & 10.0 & 7.5 & 73.0\% \\
\textbf{ASE} & 25.9 &  24.2 & 93.4\% & 25.2 & 15.9 & 63.0\% \\
\textbf{FSE} & 6.4 & 5.6 & 87.4\% & 5.4 & 2.6 & 48.1\% \\
\textbf{MSR} & 2.3 & 2.1 & 91.3\% & 4.2 & 3.4 & 80.9\% \\
\textbf{ICSME} & 3.8 & 3.1 & 81.5\% & 4.7 & 3.5 & 74.4\% \\
\textbf{SANER} & 1.4 & 1.1 & 78.5\% & 4.1 & 3.1 & 75.6\% \\
\textbf{TSE} & 2.8 & 1.9 & 67.8\% & 2.9 & 0.6 & 20.6\% \\
\textbf{EMSE} & 1.8 & 1.7 & 94.4\% & 1.2 & 0.8 & 66.6\% \\
\midrule
\midrule
\textbf{Total} & 51.8 & 46.6 & 89.9\% & 57.7 & 37.4 & 64.8\% \\
\end{tabular}
\end{table}

\begin{table}[t]
\centering
\caption{Mean interactions with GitHub repositories by declared purpose.}
\label{tab:total_interactions_Cat}
\begin{tabular}{l|cccccc}
\textbf{Categories} & \multicolumn{3}{c|}{\textbf{Pull Requests}} & \multicolumn{3}{c}{\textbf{Issues}} \\ \hline
\textbf{} & \textbf{Created} & \textbf{Closed} & \textbf{Rate} & \textbf{Created} & \textbf{Closed} & \textbf{Rate} \\
\toprule
\textbf{Tool} & 7.0& 6.3 & 89.8\% & 9.1  & 6.0  & 65.7\% \\
\textbf{Dataset} & 5.4 & 5.2 & 96.7\% & 6.1 & 3.8& 61.5\% \\
\textbf{Rep. Package} & 0.05 & 0.05 & 100.0\% & 0.0 & 0.0 & -- \\
\textbf{Tool \& Dataset} & 1.0 & 0.6  & 63.6\% & 1.9  & 1.4 & 61.5\% \\
\midrule
\midrule
\textbf{Total} & 13.4 & 12.1 & 89.6\% & 17.1 &  11.2 & 65.3\% \\
\end{tabular}
\end{table}

\subsubsection{Types and Rates of Interactions}
We first statistically analyze the most common types of public interactions with repositories on GitHub: Issues, which raise potential problems and discuss their resolution, and Pull Requests (PRs), which package proposed changes for review.
Table \ref{tab:total_interactions} shows the mean number of these interactions for each venue and in total. The results show a similar spread in popularity as before (though here too, the mean is often much higher than the median), with especially the journal venues receiving very little public interest. In practice, this is often a natural correlate with previously shown popularity metrics. 
Both types of interactions were quite common in general, with an average of ca. 8 issues and 5 PRs per repository.
However, these are yet again heavily \emph{concentrated}; only about one third of projects in our dataset received at least one interaction of either kind, and those that did frequently had either a few or hundreds such interactions in total.
Table \ref{tab:total_interactions} also offers a preview of our next subject of study: author responsiveness. While pull requests are virtually always resolved (``closed", either merged or rejected), the issue resolution rate is substantially lower in our dataset.

Table \ref{tab:total_interactions_Cat} breaks down the aggregated (mean) statistics of public interaction by repository category. Interestingly, while both tool and dataset related repositories are popular targets for contributions, the latter report the highest average number of pull request ($25.4$) and issues. Both resolve pull requests at a high rate, but are somewhat less effective at closing issues (a common pattern on Github more broadly). 


We specifically studied every \emph{first} issue of each repository to gain an even insight into the types of concerns commonly brought up when releasing a new research project. We found that these most commonly concerned \textit{installing/building} ($36.4\%$) issues, in which the developers either asked how to install the required framework and dependencies, or brought up problems with replicating the configuration, e.g., due to differences in OS. Next, $28.3\%$ related to \textit{bugs} while using the tool, such as missing files or unexpected exceptions. Not all these ``issues" were about problems: we found some 25\% simply asked for more details or information about the repository or publication, and a small number ($6.7\%$) recommended improvements or new features to add.

\summarybox{Finding \#8}{Much like popularity, repositories vary widely in public interest and engagement. Yet a fair share receive some inquiries upon launch, which often point out problems, and are sometimes constructive.}

\subsubsection{Author Responsiveness Rates to Public Interactions.}
Table \ref{tab:total_interactions} suggests that author responsiveness is quite varied, both between venues and types of interactions. We study responsiveness to issues in more detail, as these are often not resolved, focusing on participants and comments. Surprisingly, we found that the majority of issues were apparently closed without any response from the repository owner. Just $567$ issues (about one-fifth) in this group received a response from that group.
Upon further analysis, however, we found that this was an artifact of the common practice of hosting the repository on a custom account, typically named after the paper's tool, which allows one to include the repository in the paper during review time without breaking anonymity.
It was more common for an author to respond to such issues, and even maintain the repository, from their private account.
Thus, to avoid this confound and keep our analysis conservative, we only focused on the 47 repositories in which the repository owner responded to at least one issue. Among these, we found a still relatively low response rate of 41\%, though issues that do receive an owner-response are closed more often (81\%) than others.
Indeed, responding to and receiving Issues go hand-in-hand, a strong indicator that author responsiveness incentivizes future engagement.
Even then, the average time to receive a response from a repository owner was ca. $23$ days (again, some quickly while others take months or years). It was not uncommon to observe a wide spread of response rates on the same repository; popular repositories often struggle to keep up with submitted issues on GitHub.

More generally, issues varied widely in engagement: while the average issue received $2$ comments and involved 2 distinct participants (both including the initial comment), this number was heavily skewed by both one-third of issues that receive no response at all, and ca. 10\% ($219$) that received five or more comments, while $2.7\%$ triggered a discussion with ten or more responses.
We manually analyzed the issues with the most engagement. These regularly concerned discussions of bugs found by users (in $56\%$ of cases) as well as more general discussions about the projects and its results involving owners or collaborators. We note that, as may be expected, issues with a large number of comments get closed much faster than others and are very likely to include responses from repository owners.

\summarybox{Finding \#9}{Author responsiveness to issues is generally low and slow: less than half of issues receive a response from the owner, and those often take more than three weeks. More responsive authors often promote more public interaction.}

\subsubsection{Typical Responsiveness Speeds and Trends.}
\begin{figure}[!ht]
    \centering
    \vspace{-.4cm}
    \subfloat{
    {\includegraphics[width=.4\linewidth]{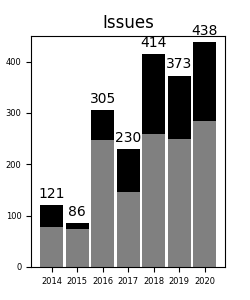}}}%
     \hskip -2ex
    \subfloat{{\includegraphics[width=.4\linewidth]{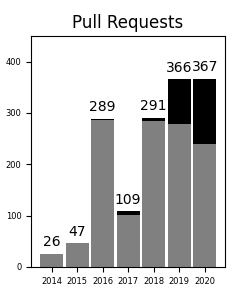} }}%
    \caption{Distribution of the number of common types of public interactions by year across all academic repositories on GitHub (\codebox{\gray{?}}: Closed \codeboxb{?}: Open).}%
    \label{fig:closed_distribution}%
\end{figure}

Figure \ref{fig:closed_distribution} shows the trends in the total number of issue \& PR events since 2014, in terms of the annual number of opened and closed items (overlayed).\footnote{Technically, some ``closed" events may belong to items ``opened" in the prior year, e.g., if one was opened in 2016 and closed it in 2017, the former observation is added to the 2016 ``opened" bar and the latter to the 2017 ``closed" bar. We count these observations separately since not all issues \& PRs are closed. This also explains why 2020 has a relatively high number of ``Opened" PRs relative to ``Closed" ones.}
The results support that pull requests are closed much more commonly than issues, and have generally become more commonplace in recent years. Issue responsiveness, meanwhile, has changed little.

\begin{figure}[!ht]
    \centering
    \includegraphics[width=0.8\linewidth,trim=40 110 40 80]{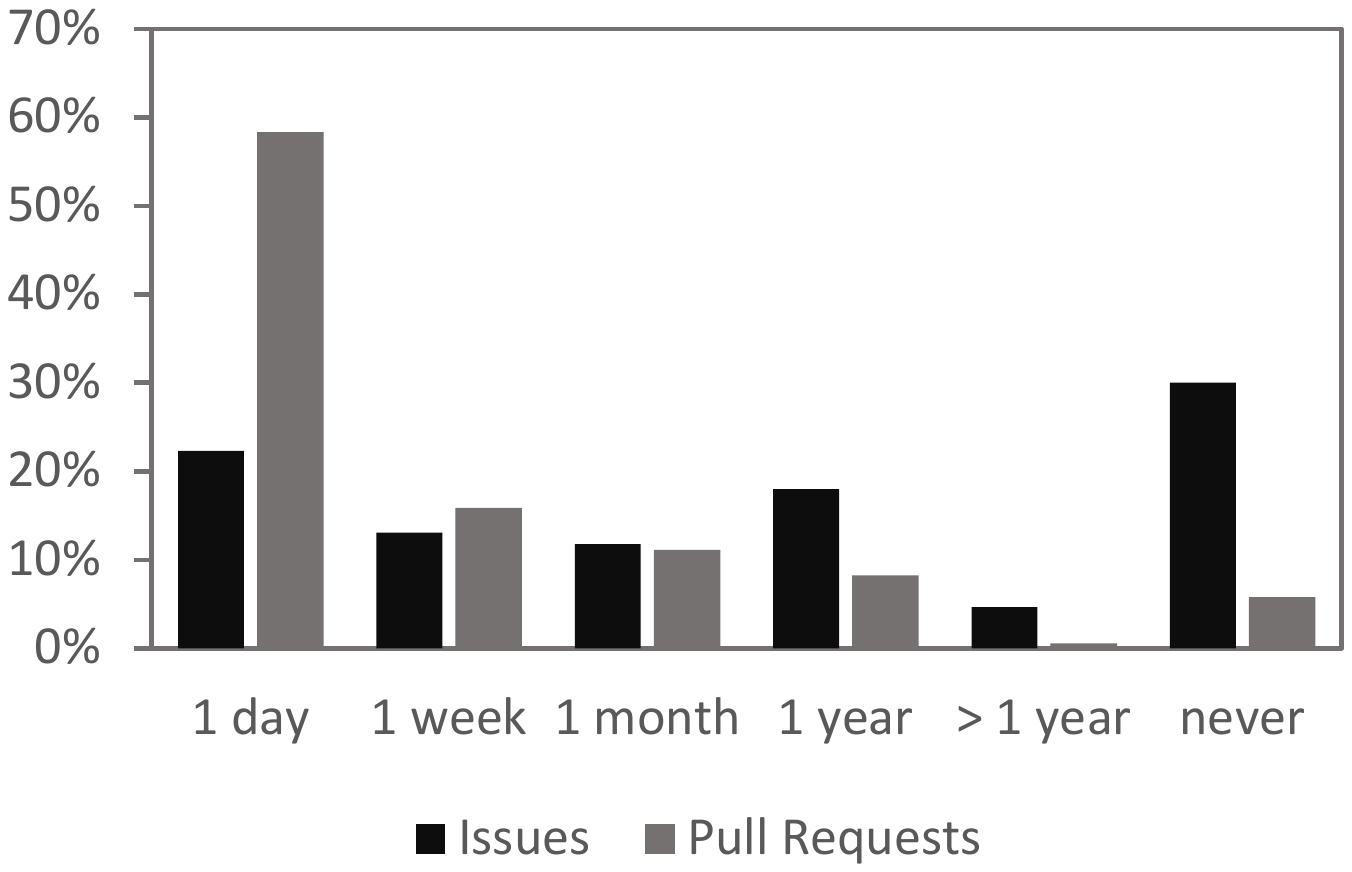}%
    \caption{Typical duration of public interactions with academic repositories, in terms of Issue and Pull Request time-to-close.}%
    \label{fig:duration_distribution}%
\end{figure}

We next analyze the \emph{average time} to close issues and pull requests in Figure \ref{fig:duration_distribution}. Here, too, the gap between Issues and PRs is apparent: PRs were approved or refused much faster, within $16$ days on average but within a single day in more than half of cases, than Issues, which are much more commonly never closed, or only after months or even years. We note that ``closing" an issue is not quite equivalent to the underlying problem being resolved, but these two tend to go hand-in-hand.
We did note a promising sign -- a marked change in the issue time-to-close in recent years: that average is $87$ days overall, but this was significantly worse in earlier years ($120$ days for issues created before $2017$) than it has been since ($67$ days).\footnote{This duration has not changed significantly for Pull Requests in recent years (15 days before 2017, 20 days since).}
Even these recent response rates, however, remain significantly worse than among other (popular) open-source projects \citep{Kikas}.

\summarybox{Finding \#10}{Contributions (in the form of Pull Requests) to academic repositories are rarer, and apparently more treasured, than concerns (Issues). The former are responded to quickly and commonly, while the latter often linger for months or years.}

\subsection{Publication \& Repository Outcomes}
\label{sec:rq3_results}

\begin{figure}[!ht]
  \centering
  \includegraphics[width=.8\linewidth]{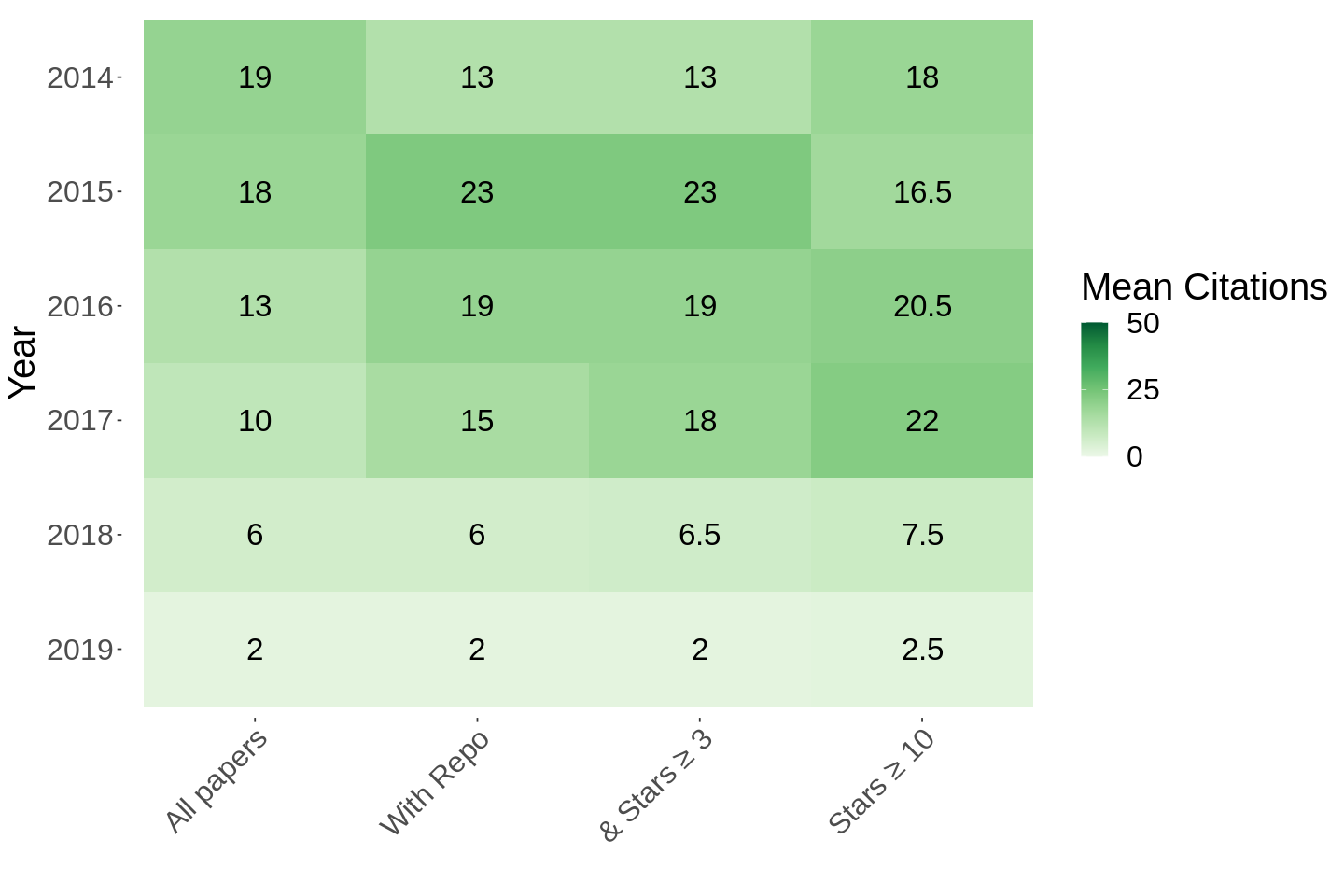}
  \caption{Median paper citations vs. repository popularity, by year, of both all papers (first column, those indexed by SemanticScholar in our venues), and of the 245 still-available repositories that specifically release a \emph{tool}, further separated by minimum star-count. Since 2016, papers with repositories were consistently cited more than those without, often increasingly so with more stars.}
  \label{fig:citations}
\end{figure}

\subsubsection{Relation to Citation Count}
Research publication success is typically measured by the total number of citations a paper receives, whereas a \gh repository's popularity is commonly indicated by its number of stargazers (stars, for short). Do these intersect for publications with repositories? We analyzed the mean citation count of all papers in our selected venues since 2014, and contrast these with those in our dataset Figure \ref{fig:citations}, using the SemanticScholar dataset to extract citation counts. Overall, recent years suggest an inversion, in which publications with repositories gained in (citation) popularity relative to the base rate -- where the former was lower in the first few years, when \gh was still relatively new, this trend has been consistently reversed since 2016.

When we further divide the latter into repositories with at least a non-trivial ($\geq 3$) and a modest ($\geq 10$) number of stars respectively, the trend is largely upwards, with more popular repositories correlating with more cited papers (Spearman's rank correlation shows $\rho \approx 0.17$, $p < 0.01$). While we cannot derive causality from this type of analysis, the correlation is encouraging: our prior findings suggest that gaining popularity on \gh requires investment (in terms of responsiveness to concerns, and providing detailed installation instructions), often continued for months or years after the publication. This analysis at least suggests that those investments tend to pay off over time. In our next findings, we further detail examples of publication-tied repositories inspiring further (and paper-citing) research.

\summarybox{Finding \#11}{While cause and causality are impossible to establish, statistics from recent years support a correlation between \gh inclusion and citation count, which is especially strong when the repository is also popular, showing that well-appreciated papers and well-liked repositories often go hand-in-hand.}

\subsubsection{Broader Impact in Open-Source Software.}
We studied the ten most \underline{forked} academic repositories to attempt to identify how and why these repositories were used. These include many repositories with hundreds of forks, like logpai/logparser ($269$ forks, at the time of this writing), cliffe/SecGen ($237$), as well as logpai/loghub ($187$), SVF-tools/SVF ($150$), sosy-lab/sv-benchmarks ($150$), honeynet/droidbot ($133$), hussien89aa/ MigrationMiner ($133$), GumTreeDiff/gumtree ($111$) and gousiosg/GitHub-mirror ($99$).
We manually check a wide range of the profiles of users that forked these repositories, to identify e.g. industry affiliations, or indicators of intent. This turned out to be a complex endeavour: user profiles are often very sparsely populated, offering few hints of the reason for their interest, and industry adoption of these tools appears not to be declared in any obvious way in most cases. We did find that a majority of these users were apparently not in academic positions (at least, not evidently so from their profile page), but were unable to establish more than a few industry affiliations, and even those are scarce evidence of industry adoption. This remains a subject for further research.

\subsubsection{Academic Impact of Publication Repositories}
Since all our studied repositories originate from academic publications, it is likely that their impact extends into subsequent academic work as well. Indeed, a large number of our original set of publications mentioned \gh repositories not owned by the publication authors. We intersected those $1,536$ links with the $342$ author-owned repositories and found a few publications that referred to published work by their \gh repositories. However, it was substantially more common for those using such tools to refer to the underlying publication -- a sign that the research accompanying these tools is well-known in the community. We discuss several prominent cases here.

\citep{Jean-Remy} published a tool called GumTree,\footnote{\url{https://github.com/GumTreeDiff/gumtree}, which computes AST-level `diff' edit scripts meant to be intuitive to developers} in ASE'2014 and made the source code publicly available in GitHub. This publication is both well-cited ($320$ citations, at the time of this writing) and its GitHub repository is very popular among developers ($128$ forks, $494$ stars). Multiple subsequent academic papers relied on GumTree, including Iterative Java Matcher, \citep{VeitFrick}
\footnote{\url{https://github.com/VeitFrick/IJM}}. Other work used GumTree as a supporting tool, such as \citep{Dongsun} and FixMiner\footnote{\url{https://github.com/SerVal-DTF/fixminer_source}} \citep{FixMiner}, to extract repair actions for use by subsequent models.

\citep{Tantithamthavorn} published a TSE paper with a repository for computing their proposed Scott-Knott Effect Size Difference (ESD),\footnote{\url{https://github.com/klainfo/ScottKnottESD}} which has found application in ranking features in defect prediction. Their method was evidently helpful to subsequent work in defect prediction. For instance, \citep{Fabio} applied their techniques in their model, and \citep{Masanari} studied the features' impact on five different validation models to compared their results.

Other publication-tied repositories serve to host benchmarks, which are often useful and can become quite popular in subsequent work. Notably, SV-benchmarks\footnote{\url{https://github.com/sosy-lab/sv-benchmarks}} is a benchmark of verification tasks that was used for experimental evaluation by both \citep{Beyer} and (a subset thereof) by \citep{Gerrard}

\summarybox{Finding \#12}{Popular repositories have the potential to greatly facilitate (and even inspire) subsequent research, which feeds back into the paper's impact.}

\section{Discussion}\label{sec:discussion}
Having established a range of distributions, statistics, and general observations, we now reflect on these to identify broader implications and recommendations. We do so primarily by focusing on various key categories of repository uses and misuses that we saw in Section \ref{sec:Results} and provide some anecdotal insights based on observations within these categories. We derive several key recommendations from these findings.

\subsection{Maintenance}
\label{sec:maintenance}
The most salient finding in this work concerns author responsiveness to public interactions, especially Issues, which play a key role on GitHub as the primary facilitators of communication between developers and maintainers.\footnote{We note that \gh recently rolled out a ``Discussions" feature, which may come to supplant Issues for this purpose.} We found that these often communicate problems with artifacts, which are naturally especially common when the code is first released. It is therefore rather troubling that so many of these never get closed in our dataset, and even more apparently fail to receive a response from a project maintainer. Manually inspecting some of these cases, we find a couple of common explanations for this summary statistic. Most obvious is a large component of author unresponsiveness. This appears to occur equally often in very busy projects, where resources are simply too scarce to address all concerns, and among ones with just one or two issues, where we found several instances of bugs or serious concerns being pointed out and apparently going entirely ignored. As Table \ref{tab:total_interactions_Cat} reflects, such repositories tend to belong to the ``replication package" category, which tend to receive just one or two issues.

Interestingly, we also find the occasional feature misuse: in some cases, most or all issues were opened by the original author(s), apparently with the intent of improving a project. Sometimes these issues were even addressed in further commits, but simply never marked resolved.
Finally, we studied some of the (majority of) repositories that never get any such Issues. Among these were many that served just to host datasets or simple scripts, but also some cases of well-documented and structured code, some quite popular. There is evidently a component of randomness between popularity, quality, and interactions that should not be overlooked.

We analyzed several repositories belonging to the two venues with remarkably low average popularity: TSE and EMSE. We found that these repositories often served as just a data store or to hold examples and scripts --, sometimes not even essential to replicating and reproducing the published work. Many of these fell into the (self-declared) replication package or dataset categories. Such repositories were found in all venues, though less singularly concentrated; these often draw just one or two issues, typically from other researchers attempting to use or replicate the data, which then tend to be ignored.
Vice versa, we found that high-interaction repositories often contained code for tools, or broadly useful datasets (e.g., benchmarks \citep{SoSy-Lab}, GitHub-mirror\footnote{\url{https://github.com/gousiosg/GitHub-mirror}}), as reflected in Table \ref{tab:author-owned_cat}.
These repositories tended to have more responsive owners, which may enable a feedback loop that draws more popularity and contributions.

This brings us to our first recommendation: \textbf{Encourage maintenance of research repositories.} While authors cannot update and improve their papers after getting published (arguably a useful limitation), this is not the case for code, which provides a natural avenue to evolve and clarify a paper's contributions in response to public interest. Yet many of these repositories were never updated since being published to GitHub, often despite requests and concerns that ought to prompt it to. It is not clear why some authors choose to publish on \gh (over a more archival solution like Zenodo) given this apparent disinterest in maintenance. More research is needed to understand this phenomenon, but in the meantime, our findings suggest that it would be beneficial for our community to clearly articulate the benefit and merit of repository upkeep after paper acceptance. This merit is real: we find a sound correlation between well-cited papers and popular repositories. This should include communicating these expectations when reviewing work with repositories, but also rewarding those whose repositories had a meaningful impact -- much like we reward impactful papers.

\subsection{Documentation} Many issues in our dataset concerned questions about configuring, installing and building the research code. We found that SE research repositories often came with rather poor documentation, and lacked explanations and visualizations of datasets and benchmarks. We note that, while some repositories lacked any instructions, many did indeed include a README file with instructions; unfortunately, this is rarely sufficient and many issues concerned flaws in these steps, or the assumptions they made. 

Thus for our second recommendation: \textbf{provide more thorough documentation}. Poor documentation hampers or entirely prevents others from using code. Many conferences now include an artifact evaluation track, the entire purpose of which is to check and verify (and often correct, interactively over many weeks), the quality of research artifacts. Many of the papers in our dataset have evidently not gone through this process, as issues pointing out problems with the setup were some of the most common -- and frequently unaddressed. This is an unsustainable environment for research quality assurance and replicability. As a community, we should both develop, and enforce  guidelines to improve documentation quality. Consideration should be given to the common practice of omitting repository links during the review period to preserve anonymity and adding them in the camera-ready. A guided process may be more appropriate here, akin to a conditional accept.

\subsection{Broader Connections to the GitHub Ecosystem}
\label{sec:broader}
Finally, our analysis shows a steep increase in references to \gh URLs in general, many of which were used as examples or dependencies in the papers' design. In this section, we include a preliminary analysis of these ``other links" and the role they play. More generally, these many connections present a new kind of complication, not so much in existence as in scale, related to dependability. First, even among the links tied directly to publication repositories, 4 of the 342 were no longer available -- a small enough number for now, but also an alarming reminder that \gh is not an archival resource. Beyond that, as more and more \gh links and projects factor into papers and paper-repositories, we are likely to see increasingly many ``secondary" replicability failures: examples cited in papers disappearing, and projects becoming impossible to build. Thus, we strongly recommend \textbf{Hosting source code and dataset on archival services:} paper-related data, of any kind, should be permanently archived in an appropriate resource (perhaps more than one). This is not at all to the exclusion of \gh repositories; services like Zenodo can track and archive repositories at every release. We found remarkably few uses of this feature, however. Its merit should be more broadly advocated.

We briefly detail some of the usages of non-author-owned \gh links in our dataset.
\paragraph{1) Used for implementation:} where the authors used the repository in their research project, such as extending the source code or using a \gh hosted dataset. We found $458$ such uses in our publications. Several of these were completely reliant on that repository. This was common when using ML frameworks, like Keras \citep{keras}, an open-source software library to implement artificial neural networks, mentioned in four publications \citep{Santos, Zhao, Tian, Du}. Similarly, WALA \citep{WALA}, a Java library for static and dynamic program analysis, was used to heavily by \citep{Zhi, Wang, Luo}, as was cloc \citep{cloc}, to count the lines of source code \citep{Mamun1, Gadient, Osman}. This does not solely apply to tools: some benchmark repositories such as SV-Benchmark \citep{SoSy-Lab} were also broadly impactful.

\paragraph{2) Motivation:} where the authors use a \gh project as a motivating example for their work. The vast majority of these were mentioned in the introduction section.
This is especially common when the publication centers around studying GitHub-specific phenomena, such as issues or proposing new tools for the GitHub community \citep{Zou, Yun}. We found $119$ such links to projects alone -- hundreds more referenced specific issues and pull request for similar reasons.

\paragraph{3) Example:} where the authors refer to GitHub as an example of a task, problem or a feature of GitHub such as issues and bugs.
This is especially common, with $901$ mentions of repositories in our dataset, which includes comprehensive empirical studies (much like this one) and surveys. GitHub repositories are popular among these because of the large number of publicly available projects. For instance, \citep{webview} proposed automated testing techniques and conducted an empirical study on bug testing by using $31$ GitHub repositories. DroidLeaks \citep{Liu} is a comprehensive study that collected $124,215$ code revisions on $34$ GitHub repositories to create a bug database. This is just the tip of the iceberg: many papers use hundreds or thousands of \gh repositories in their datasets even if they are not explicitly mentioned. No doubt many of those repositories are already no longer available.

\paragraph{4) Related work:} where the authors discuss the related projects to their proposed tool or benchmark. We find $55$ such repositories. For instance, \citep{Cliffe} proposes \citep{SecGen} and discuss its advantages over 5 existing GitHub repositories, not all of which from research.

These categories can and do intersect. The most commonly cited GitHub repository overall was Google's Guava \citep{Guava}, which is used five times as an example, three times as motivation and twice for implementation.

\section{Related Work}\label{sec:Related}
The intersection of research code and publications is well studied.
\citep{Supatsara} studied links to research publications from GitHub repositories, effectively the reverse of ours, across $20$ thousand projects. Most of these publications were in the Machine Learning community. We note that many repositories in our dataset did not (clearly) refer to the underlying publication and would have gone overlooked by such an analysis. Other analyses are complementary to ours, such as \citep{Bangerth}, who studied how scientific software projects can be successful.
Academics have long been known to play an important role in open-source development.
Many popular GitHub repositories related to Machine Learning were created by academic researchers \citep{Braiek}. \citep{Milewicz} find that, among scientific GitHub repositories, most contributions came from academic researchers and students.

Others explicitly analyzed research artifacts published with papers.
\citep{Timperley} is especially relevant here, as they also focused on papers in SE. Among others, they find many artifacts hosted on other platforms, such as Google Drive, which is to be expected when focusing specifically on replication artifacts. Their findings correspondingly suggest that researchers share their artifacts primarily to facilitate replicating, reproducing or building on their work.
Storer \etal~ \citep{Storer} studied the state of software in academic research, but  not in the context of open-source interactions.
\citep{Inokuchi} investigate the contribution of Software Engineering researchers to software development by analyzing publication links in source code comments, mostly on GitHub.

Several studies relate closely to some of our findings. On (sustained) maintenance and popularity, \citep{Mockus} showed the importance of long time contributors to the success open software projects.
\citep{Collberg, Collberg2} analyzed the source code of $601$ papers and its efforts to run and build the code. A non-trivial portion of these repositories did not run or required extra work to run it.
Numerous studies have aimed to understand and characterize publication impact in terms of citations \citep{Donner,Redner,Bornmann}, as well as popularity on GitHub, where e.g. \citep{Borges} found a strong correlation between stars and forks. In our studies, we connect these two by finding a correlation between GitHub repository popularity (star and fork) and publication citations.
\citep{Weber} also studied popularity of GitHub repositories, specifically focusing on Python Machine Learning GitHub repositories.
\citep{Marks} also studied the lifetime of issues and found substantial variation in time to resolve bugs in Mozilla and Eclipse.

\section{Threats to Validity}\label{sec:threatstovalidity}
\paragraph{Construct Validity:} numerous research artifacts have been created to archive code. We only consider GitHub because it is the \emph{largest, social} open-source platform, used by the overwhelming majority of software developers. These traits were more important to us than expanding our dataset, as our study specifically focuses on understanding the social implications of publishing research code on a platformed designed to enable public interactions.
Similarly, our study is scoped to the software engineering research community alone, as it (all-but) uniquely encompasses researchers that both frequently study social coding and engage in it. Future work may expand our analysis to other domains that commonly use GitHub, such as Machine Learning and Human Computer Interaction.
Similarly, we focused only on a selection of top venues of software engineering. This selection encompassed the majority of highly rated venues, including most A and A+ rated conferences and the two top journals in our field. It is possible that repositories associated with lower-rated publications have significantly different characteristics; future work may choose to study those.

\paragraph{External validity:} Our dataset was limited by public paper availability to $10,034$ publications. Since the practice of providing a copy of the paper online has recently become more popular, we naturally expect the numbers in our paper to be skewed by this trend and therefore inexact. It may be a mitigating factor that most GitHub repositories were also published recently; regardless, we argue that our general findings, such as differences between Issues and Pull Requests, are unlikely to be affected by these criteria.

\paragraph{Internal validity:} We manually annotated the ca. 3.5K \gh links in our dataset. Annotating these links often required visiting the URL (unless they obviously referenced a third-party repository, or other link GitHub link) and matching the GitHub repository owner to the publication's author. This is technically a noisy process, though in practice it was virtually always very clear whether a repository was tied to the publication.

\section{Conclusion}\label{sec:conclusion}
We present an empirical study of software engineering researchers' use of \gh to release code associated with publications. \gh is a social platform, and repository owners are expected to engage with the public. Our findings suggest that this rarely works out in practice: only some repositories gain popularity and attract public comments (not all negative), and those that do often neglect, or take weeks, to respond to them, even if they point out real concerns with the artifact. The few projects that do remain active are often successful, sometimes tremendously so. These more successful cases additionally correlate strongly in popularity with the citation count of the related paper, and indeed often clearly enabled future work through their repositories. Our findings suggest the need for reconsidering both the merit of, and incentives for, including repositories with published work.

\section*{Data availability}
The datasets collected and analyzed during this study are available in the GitHub repository, accessible at https://github.com/Kamel773/paper-with-GitHub-link/.


\bibliographystyle{IEEEtran}
\bibliography{references}

\end{document}